\documentclass{aa}

\usepackage{graphicx}
\usepackage{natbib}
\usepackage{txfonts}

\begin{document}

\title{Detailed AGB evolutionary models and near infrared colours of intermediate-age 
stellar populations: Tests on star clusters}
\author{M. Salaris\inst{1, 2} \and A. Weiss\inst{2} 
        \and L.P. Cassar\`{a}\inst{3,4} \and L. Piovan\inst{3}
        \and C. Chiosi\inst{3}}
\institute{Astrophysics Research Institute, Liverpool John Moores
  University, IC2, Liverpool Science Park, 146 Brownlow Hill, 
 Liverpool, L3 5RF, UK \\
  \email{M.Salaris@ljmu.ac.uk} 
  \and Max-Planck-Institut f\"ur Astrophysik,
  Karl-Schwarzschild-Str. 1, Garching bei M\"unchen, Germany \\
  \email{aweiss@mpa-garching.mpg.de} 
  \and Department of Physics and Astronomy, University of Padova,
       Via Marzolo 8-I, 35131, Padova, Italy
  \and INAF-IASF Milano, Via E. Bassini 15, 20133 Milano, Italy
          }
\date{Received, accepted}

\abstract{We investigate the influence of Asymptotic
  Giant Branch stars on integrated colours of star clusters of ages
  between $\sim$100~Myr and a few gigayears, and composition typical for the
  Magellanic Clouds. We use state-of-the-art stellar evolution 
  models that cover the full thermal pulse phase, 
  and take into account the influence of dusty envelopes on the
  emerging spectra. We present
  an alternative approach to the usual isochrone method, and compute
  integrated fluxes and colours using a Monte Carlo technique that enables us to 
  take into account statistical fluctuations due 
  to the typical small number of cluster stars. 
  We demonstrate
  how the statistical variations in the number of Asymptotic Giant
  Branch stars and the temperature and luminosity variations during
  thermal pulses fundmentally limit the accuracy of the comparison (and 
  calibration, for population synthesis models that require a calibration of the 
  Asymptotic Giant Branch contribution to the total luminosity) with 
  star cluster integrated photometries.
  When compared to observed integrated
  colours of individual and stacked clusters in the Magellanic Clouds, our predictions  
  match well most of the observations, when statistical fluctuations are taken into account, alhough 
  there are discrepancies in narrow age ranges with some (but not all) set of observations.}

\authorrunning{M. Salaris et al.}
\titlerunning{New AGB models and near infrared integrated colours}
\maketitle

\section{Introduction}
\label{s:intro}

Stars of moderate mass, between approximately 1 and 8~$M_\odot$, spend 
less than 1 percent of their lifetime in the phase of double-shell (H-
and He-) burning, generally called the Asymptotic Giant Branch (AGB) phase,
after exhaustion of central helium-burning and before turning into
white dwarfs. For an even shorter fraction of their nuclear life, of
order $10^{-3}$, they experience the thermal pulses (TP-AGB phase),
during which temperature and brightness vary significantly on
timescales of a few thousand years. On the other hand, in these
evolutionary stages they are much brighter than they were on the
main-sequence (MS): stars of intermediate mass ($\approx 2 - 2.3 \cdots 6 - 8\,
M_\sun$) that do not develope electron degenerate He-cores after the MS  
(the exact mass range depending on the initial chemical composition 
and treatment of core convection) have a bolometric luminosity larger 
by a factor of up to 100 compared to the MS turn-off.
For low-mass stars the difference is even
larger. However, for the latter the difference between the TP-AGB
luminosity and that on the tip of the Red Giant Branch (RGB) is closer
to a factor of only 10. Since AGB stars are also much cooler than
MS stars of comparable mass, they may completely dominate
the IR integrated light of a stellar population. 
The integrated red magnitudes and colours of a single-age stellar 
population will be dominated by AGB-stars at an epoch
when intermediate mass stars have reached this evolutionary stage. At
earlier times massive stars dominate, but they contribute much less in 
the red, because they increase their brightness relatively moderately
in post-MS phases and spend a much larger fraction in the blue, anyway. At
later times low-mass stars reach the AGB, but here the longer-lived RGB stars 
dominate the IR output, and thus the importance of the AGB stars is
reduced. The age range, during which intermediate-mass AGB stars are
significant for the integrated spectrum, is therefore between $\sim$100~Myr and 
$\sim$1Gyr, varying somewhat with metallicity. This has been demonstrated
frequently \citep{lm:00,bch:2003,mar:05,zlhzk:2013,ip:2013} in models
of population synthesis \citep[see, for 
    example,][for a recent review]{mgba:2010}.

In a broader context, the correct prediction of the IR integrated flux of a stellar population 
is crucial, for example, to disentangle the age-metallicity degeneracy in integrated colours of unresolved stellar 
populations \citep[see, e.g.,][]{anders, j:06}, and determine stellar masses of high-redshift objects \citep[see, e.g.][]{mar:05}.

New fully evolutionary AGB stellar models for
moderate-mass stars, which treat the
effects of nucleosynthesis and mixing of carbon on stellar effective temperatures 
and mass loss consistently, are now available \citep[][hereinafter WF09]{wf:2009}, 
and in a recent paper \citep[Paper~I]{cp:2013} we have
presented new population synthesis models where the contribution of AGB-stars to integrated
spectra and colours made use of these AGB calculations\footnote{Both WF09 AGB evolutionary models and 
the population synthesis models of Paper~I are available upon request}. 
In the same Paper~I, the reprocessing of stellar
light by the dusty circumstellar shell, as obtained from the stellar
models, was taken into account by a library of spectral energy
distribution (SED) models for such cases. In doing so,
we could investigate the role of dust in models of
population synthesis for intermediate age populations 
\citep[see, e.g,][for earlier works on this subject]{ptc:2003, mar:08}.
Paper~I was the first attempt, to date, to include fully evolutionary AGB 
calculations in the predictions of integrated spectra of stellar populations. 
The standard approach is to include instead simplified treatments of the AGB phase \citep{mar:05}, 
or synthetic AGB calculations \citep[see, e.g.,][]{mar:08}.

However, in several aspects Paper~I remained restricted to conventional assumptions and procedures. 
First, a preexisting widely used set of isochrones \citep{bbcfn:94} is employed for the pre-AGB phase, 
and the AGB part is {\sl attached} after appropriate $T_\mathrm{eff}$ and bolometric luminosity shifts, to ensure continuity.
This procedure destroys in principle the self-consistency of the AGB calculations; just as an example,  
the mass loss rates depend on the model luminosity and $T_\mathrm{eff}$, and shifting the tracks would require 
rates different from the ones actually used, thus a different AGB evolution. Besides this 
intrinsic inconsistency, the 
contribution of the AGB (calculated from the shifted tracks) 
to the integrated flux can be potentially significantly different from what 
predicted by the {\sl original} AGB calculations.
We notice here that 
shifts in ${T_\mathrm{eff}}$ and luminosity of 
existing AGB calculations to match pre-AGB models and/or satisfy empirical constraints,  
are adopted  also in other population synthesis models, like the Flexible Stellar Population Synthesis model by \citet{conroy:10}.

Second, the TP-AGB portion of the tracks is {\sl pre-smoothed} to eliminate the complicated and irregular 
loops in the ${T_\mathrm{eff}}$-$L$ plane, 
and allow a simple interpolation when calculating the full isochrones. This is similar to reducing  
the TP-AGB phase to the evolution during quiescent H-shell burning 
as usually done when synthetic AGB models \citep[that nowadays can also take into account the detailed varition 
of luminosity and sometimes also $T_\mathrm{eff}$ along the pulse cycles, see i.e.][]{itk04, mg07} 
are implemented in the calculation of integrated spectra of stellar populations \citep[see][for an example]{mar:08},  
although the (generally small) contribution of the He-shell burning on the timescales is retained. 

Third, the distribution of stars along the AGB is computed according to an analytical
integration of the initial mass function (IMF). This is certainly appropriate (in the approximation 
of {\sl smooth} TP-AGB tracks) for well populated massive galaxies, e.g., in case of a well sampled 
AGB sequence, but it is 
well known that for stellar
clusters and low-mass galaxies the number of AGB stars can be so low that statistical fluctuations 
of AGB-dominated magnitudes become important \citep{cbb:1988,fmb:1990,sf:1997,b:02,bch:2003,ko:13}. 
This is relevant when taking into account that tests and calibrations of AGB calculations 
(evolutionary or synthetic) are typically performed on star clusters of the Magellanic Clouds.

In this study we investigate the effects of 
these three ingredients of the standard approach to include the AGB phase 
in stellar population synthesis models followed in Paper~I. To this purpose we 
extend WF09 calculations to models with masses below 1$M_{\odot}$ (the lower 
limit of WF09 computations), to calculate self-consistent isochrones for the pre-AGB evolution.
We also provide extended tests of the 
accuracy of integrated magnitudes obtained from WF09 calculations  --using several sets of data suitable for 
testing predicted integrated 
magnitudes/colours that are dominated by AGB stars-- and study the differences with selected integrated colours 
predicted in Paper~I  
and by other population synthesis models in the literature.  
Our approach is the following. We use the AGB evolutionary
tracks and all spectral libraries of Paper~I, restricted, however, to
two metallicities, $Z=0.004$ and 0.008, appropriate for most of the
SMC and LMC clusters, that are used to
test/calibrate AGB models. Section~\ref{s:models} presents a brief discussion of 
the models, introduces the extension of 
WF09 calculations to low-mass stars, and describes a new
Monte Carlo (MC) based approach to calculate the resulting integrated photometric
properties, taking into account the oscillatory behaviour of stellar
parameters during the TPs.
Section~\ref{s:prediction} discusses 
the effect of statistical fluctuations on magnitudes/colours 
dominated by the AGB contribution, and 
evaluates the effect on IR integrated magnitudes 
of smoothing the TP-AGB tracks and shifting AGB models to match 
different sets of pre-AGB isochrones as in Paper~I. We also compare 
predictions for selected colours with independent models in the literature. 
Section~\ref{s:calibrations} compares our predictions 
with several observational data, taking into account the effect of statistical fluctuations, 
followed by Sect.~\ref{s:conclusion} that closes the paper with our conclusions.
 
\section{Models}
\label{s:models}

We discuss briefly here the libraries of stellar models and
spectra employed in our analysis. 
 
\subsection{Stellar models and tracks}
\label{s:tracks}

We used the same stellar models and evolutionary tracks for stars
between 1.0 and $6.0\, M_\sun$ as in Paper~I. Details about the 
physical assumptions and the properties of these models can be found
in \cite{kitsikis:2008} and WF09. Here we only
recall that in addition to up-to-date nuclear reaction rates and 
equation of state, tables of Rosseland mean opacities for the full
temperature and density range were used that
take into account not only changes in hydrogen, helium, and overall
metallicity, but also the individual variations of carbon and oxygen. 
The abundances of these
elements change in the stellar envelope and atmosphere as a result of
the dredge-ups (the third dredge-up 
is achieved in the stellar model calculations  
by including overshooting from the Schwarzschild boundaries of the
convective regions) and, \lq{when efficient}\rq, the hot-bottom burning during the TP-AGB phase.
The opacity tables were obtained from the OPAL
website\footnote[2]{http://physci.llnl.gov/Research/OPAL}
\citep{ir:1996} for
high temperatures, and were prepared specifically for low temperatures using the
code by \citet{fa:2005}, that considers both molecular and
dust opacities. Overshooting was treated with a diffusion 
formalism as described in \cite{wsch:2008}, where also other details
about the stellar evolution code can be found. 

\begin{table*}
\caption{Stellar lifetimes (in Myr) of the models used in this paper. Columns
  2--5 and 7--10 give the duration of
 the central hydrogen burning, RGB, core-He burning, and early
AGB phases, respectively, for the two chemical compositions. The last columns of
each metallicity group summarize the age $\tau$ at the beginning 
of the TP-AGB.}
\label{t:1}
\centering
\begin{tabular}{c | r r r r r| r r r r r}   
\hline\hline
& \multicolumn{5}{c|}{$Z=0.004$} & \multicolumn{5}{c}{$Z=0.008$} \cr
\hline
$M_\mathrm{ZAMS}/M_\sun$ & 
$t_\mathrm{MS}$~~ & $t_\mathrm{RGB}$~~ &
$t_\mathrm{He}$~~ & $t_\mathrm{EAGB}$ & $\tau$~~~~ &
$t_\mathrm{MS}$~~ & $t_\mathrm{RGB}$~~ &
$t_\mathrm{He}$~~ & $t_\mathrm{EAGB}$ & $\tau$~~~~ \cr
\hline 
1.0 & 5771.4 & 2050.5 &  92.85 & 10.97 & 7925.7& 6751.9 & 2521.0 & 104.48 &
10.37 & 9387.8\\
1.2 & 3141.2 & 966.74 &  94.45 &  9.32 & 4211.7& 3600.5 & 1182.0 &  97.96 &
10.47 & 4890.9\\
1.5 & 1847.2 & 164.02 &  94.34 &  9.25 & 2114.8& 2097.9 & 193.39 &
99.37 &  9.81 & 2400.5\\
1.6 & 1525.1 & 117.01 & 103.22 &  9.08 & 1756.6& 1725.1 & 137.98 &
100.87 & 11.36 & 1975.3\\
1.8 & 1088.4 &  65.25 & 131.73 & 11.28 & 1296.7& 1223.8 &  78.85 &
128.10 & 11.80 & 1442.6\\
2.0 & 812.67 &  27.46 & 199.22 & 11.36 & 1050.7& 900.54 &  32.11 &
236.41 & 14.57 & 1183.6\\
2.6 & 410.03 &   8.59 &  82.06 &  4.07 &  505.1& 448.03 &   9.96 &
106.62 &  6.74 &  571.4\\
3.0 & 286.74 &   5.23 &  50.58 &  2.34 &  344.9& 309.24 &   5.93 &
60.20 &  3.71 &  379.1\\
4.0 & 145.00 &   2.12 &  20.48 &  1.15 &  168.8& 152.27 &   2.33 &
23.40 &  1.25 &  179.3\\
5.0 &  88.23 &   1.15 &   11.4 &  0.59 &  101.4&  90.83 &   1.24 &
12.18 &  0.69 &  104.9\\
6.0 &  60.07 &   0.71 &   7.54 &  0.34 &   68.6&  60.98 &   0.75 &
7.75 &  0.41 &   69.9\\
\hline 
\end{tabular}
\end{table*}

As a consequence of carbon-enrichment, the envelope opacities tend to
increase, leading to lower effective temperatures, which result in an
increase of mass loss. This was accounted for using
theoretical mass loss rates for for carbon-rich  
envelope compositions \citep{wachter:2002}, and empirical rates 
for oxygen-rich compositions \citep{vanLoon:2005}. 
The result is a self-consistent treatment of the
effects of the third dredge-up and a superwind-like shedding of the
stellar envelopes, that terminates the AGB and leads to the
post-AGB phase. Such treatment has by now become standard, and
similar models have meanwhile been published by \cite{vm:2010} and
\cite{kcs:2010}. 

We have used here only two out of the 10 chemical mixtures 
for which WF09 provide evolutionary tracks,  
with compositions characteristic for the
LMC and SMC: $Z=0.008$ ([Fe/H]=$-$0.35) and 0.004 ([Fe/H]=$-$0.66), 
both with initial solar metal ratios, our solar reference being
\cite{gn:93}. Table~\ref{t:1} summarizes the lifetimes of all models in
various evolutionary stages, as well as the age when they reach the
TP-AGB phase. This age will become important in the context of our
isochrone construction in Sect.~\ref{s:spectra}. To compute integrated
fluxes, we extended this 
grid with additional models for masses $0.5 \cdots 1.0\, M_\odot$
in steps of $0.1\, M_\sun$, but only until the end of the RGB. 
As explained in the next section, in contrast with the procedure in Paper~I, 
where the AGB-tracks were merged with the pre-existing and independent library of
\cite{bbcfn:94}, in the present paper we use exclusively these
tracks, all being computed with the same version of the Garching
Stellar Evolution code \citep{wsch:2008}. This
implies that we can consider the effect of AGB stars on integrated
colours only for ages within the range of $\tau$ given in
Table~\ref{t:1}, i.e.\ between approximately 60~Myr and 4-5~Gyr (log(t) between $\sim$8.8 
and 9.6-9.7).

\subsection{Library of dusty spectra}
\label{s:library}

For any set of stellar parameters (photospheric composition,
$T_\mathrm{eff}$, $\log g$, $\dot{M}$) the emerging spectrum was  
calculated in two steps: First, a stellar spectrum was 
constructed using a number of published libraries appropriate for the
various evolutionary stages and $T_\mathrm{eff}$-ranges
\citep[see Paper~I for
  details]{bbws:1989,bbs:1991,fptww:1994,ah:1995,lcb:1998}. 
The effect of the new AGB models on dust-free stellar spectra
for ``simple stellar populations'' (SSPs) was presented in Paper~I,
Sect.~5.1. In general, colours are redder compared to models that do
not take into account the enrichment of the convective envelope with
carbon, due to the lower effective 
temperatures of the models. Also the UV-flux can be increased
because of changes in the post-AGB stellar mass and transition times
compared to older AGB calculations.

In the second step, the stellar spectrum was mapped into an observed
spectrum by calculating for the AGB models 
the effect of crossing the surrounding dusty envelopes. 
This was done by computing the radiative transfer using a library
of pre-compiled spherical circumstellar envelopes. The library uses as
input parameters $L$, $T_\mathrm{eff}$, $\dot{M}$, the C/O-ratio, and
the remaining overall surface metallicity, from which, following
Eq.~(14) of Paper~I and the details given in Sect.~6 of that paper,
the wavelength-dependent optical depth $\tau_\lambda$, the dust
composition and the extinction coefficients are calculated. This
resulted in the spectral energy distribution after crossing the
shell. Details of this approach, based on \cite{ei:2001},
were developed in \cite{ptc:2003}.  For the computation of the
dusty-envelope library a modified version of the radiative transfer
code DUSTY \cite[v.~2.06]{dusty:1997} was used. More details as well
as a discussion of the theoretical spectra of dust-enshrouded O- and
C-stars, and their effect on on colour-magnitude-diagrams and 
integrated colours of stellar populations are given in Paper~I.
In the rest of the paper -- unless otherwise specified -- we will make
use of results  
obtained employing AGB spectra with dust. 
 
\subsection{Calculation of isochrones and integrated spectra}
\label{s:spectra}

We followed an alternative approach to calculate integrated
spectra, magnitudes and colours of SSPs,
that include the full TP-AGB phase, compared to the standard one,
used, for example, in Paper~I. This standard method envisages first the 
computation of an isochrone for a given age and initial chemical
composition, that covers all evolutionary stages from the MS to the
end of the AGB phase. Then a spectrum is assigned to each point along
the isochrone, making use of model atmosphere calculations (or
empirical spectral libraries).  These individual spectra are then
summed up by weighting the contribution of the individual points along
the isochrone -- each point corresponds to a value of the initial
stellar mass evolving at that stage -- according to the chosen IMF.

\begin{figure}
\centering
\includegraphics[width=\columnwidth]{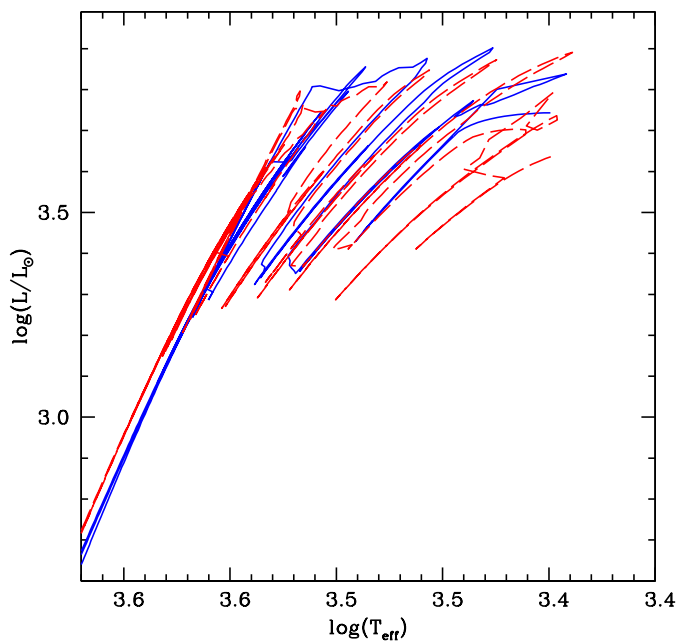}
\caption{HRD of the AGB phase of the Z=0.008,
  $1.6\, M_{\odot}$ (solid line) and $1.8\, M_{\odot}$
  (dashed line) models, respectively.
  }
\label{f:HRDagb}
\end{figure}

\begin{figure}
\centering
\includegraphics[width=\columnwidth]{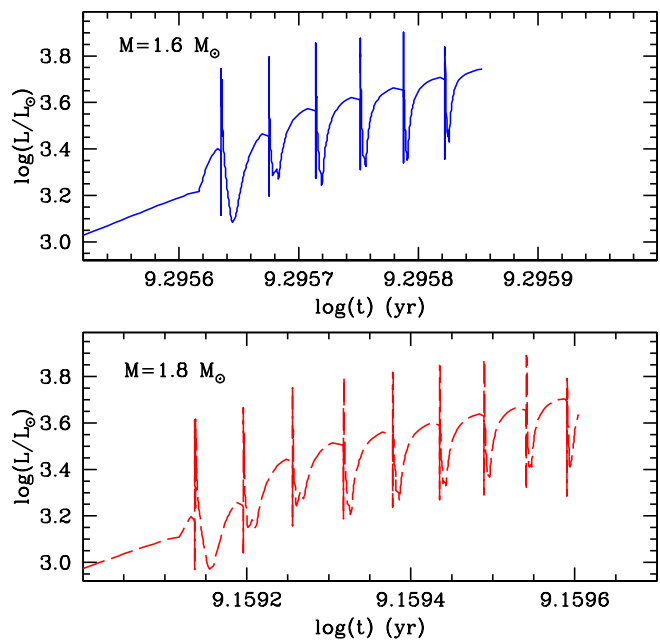}
\caption{Time evolution of the bolometric luminosity of   
the Z=0.008, $1.6\, M_{\odot}$ (solid line) and $1.8\, M_{\odot}$
  (dashed line) AGB tracks displayed in Fig.~\ref{f:HRDagb}.}
\label{Levol}
\end{figure}

The calculation of TP-AGB isochrones are however complicated by the following two 
issues:

\begin{itemize}
\item{The extremely short evolutionary timescales along the AGB (and in
particular along the TP-AGB) phase, negligible compared to the
lifetime until the end of central He burning, and comparable
  to the duration of the star formation episode. Any given isochrone
  age is therefore an idealization, and stars on the TP-AGB can be in
  any phase of pulse episodes.}

\item{The very
irregular shape of the AGB tracks in the HRD, that 
can change substantially for small variations of the
initial mass, as shown by the Hertzsprung-Russell diagram (HRD) of
Fig.~\ref{f:HRDagb}. We emphasize that the irregularity in the
  tracks shown arise mainly from that in the evolution of $T_{\rm
    eff}$, resulting from the complex interplay between a changing
  surface composition, opacities, mass-loss, and induced structural
  changes. Luminosities, in contrast, behave much more regular, as shown in Fig.~\ref{Levol}.
  Notice how both models (1.6 and 1.8$M_{\odot}$, Z=0.008) do reach similar minimum $T_{\rm
    eff}$ along the AGB. The detailed $T_{\rm
    eff}$ evolution is the result of the complicated interplay between 
   opacities, surface C/O ratio, initial mass and mass loss, but in general 
   the two tracks attain similar $T_{\rm eff}$ for similar actual mass and surface C/O ratio.  
   When the C/O ratio reaches values around 2, the opacity becomes less sensitive to these abundances 
   and at the end of the TP-AGB phase --both models having almost the same actual mass-- 
   in spite of different surface C/O ratios ($\sim$2.5 vs $\sim$ 2.0, the larger value for the model with initial mass equal to 
   1.6$M_{\odot}$) the final $T_{\rm eff}$ is essentially the same.
}
\end{itemize}

To avoid potential spurious effects due to interpolations amongst
the AGB phase of our adopted model grid, we proceeded as follows:

\begin{itemize}

\item{For a fixed initial chemical composition and age, we 
  calculated isochrones from the MS to the end of the early-AGB
  (E-AGB) using the methods described in \cite{prather:76},
  \cite{bv:92}, and \cite{basti}.  Each isochrone is made of 1300
  points. The E-AGB phase comprises 200 points.  Isochrones were
  calculated for ages equal to the lifetime at the beginning of the
  AGB phase of the set of computed AGB tracks ($\tau$ in
  Table~\ref{t:1}).}


\item{We took the evolutionary track of the mass whose
  lifetime at the beginning of the AGB phase corresponds to the age of
  the corresponding isochrone, as being representative of the isochrone AGB.}

\end{itemize} 

We call this composite sequence the {\sl evolutionary HRD} of an SSP,
and have therefore calculated {\sl evolutionary HRDs} only for ages
corresponding to the TP-AGB age of the masses in our grid.  We did not
consider intermediate values of the age, because this requires an
interpolation in mass amongst TP-AGB tracks. 
In principle 
this interpolation should be avoided because of the extremely short
timescales and the wide loops in the HRD during the TP-AGB phase, with
no regular pattern at varying mass -- especially for the lower masses
-- that would potentially cause a large uncertainty in the HRD evolution of the
interpolated track. This becomes evident when considering the
underlying idea of {\em equivalent evolutionary points}
\citep{prather:76, basti}  for the usual isochrone strategy. 

\begin{figure}
\centering
\includegraphics[width=0.8\columnwidth]{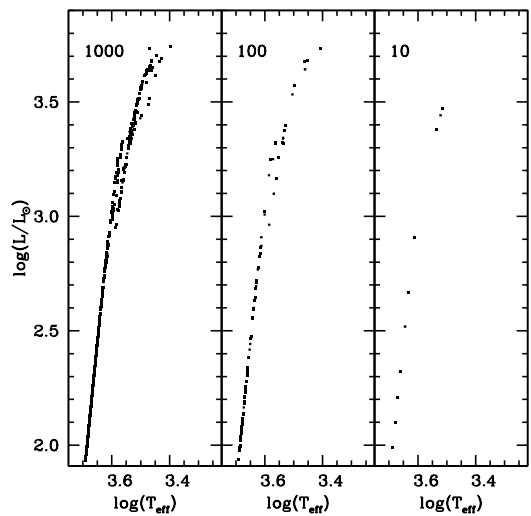}
\caption{Examples of Monte Carlo simulations of the AGB phase for Z=0.008,
  t=1.19~Gyr (log(t)=9.08), and the labelled numbers of AGB stars. Each panel
    shows one out of 100 MC realization of the respective number of
    AGB stars.
}
\label{f:cmd}
\end{figure}

Given the {\sl hybrid} nature of our {\sl evolutionary HRDs}, and the
aim of determining the fluctuations of the integrated magnitudes
sensitive to AGB stars, we had to resort to a non-standard way to
determine the integrated flux of each SSP, that makes use of both MC
simulations and analytical flux integrations:

\begin{itemize}

\item{At a given chemical composition, for each age we calculated
  synthetic HRDs of both E-AGB (the reason for including also the E-AGB
  phase is explained below) and TP-AGB phase.  Given that these phases 
  are represented by the evolutionary track of a single mass, the
  synthetic HRD has been calculated by drawing randomly age values
  within the range covered by the full AGB phase, with a uniform
  probability distribution.  Interpolation in age along the track
  provided the HRD location of a representative synthetic AGB object.
  For each age and chemical composition we
  computed five sets of 100 realizations of the representative AGB, for
  numbers of stars equal to 10, 30, 100, 300 and 1000, respectively.
  Examples of MC HRDs of the AGB phase 
  are shown in Fig.~\ref{f:cmd}.  For each realization and each number
  of TP-AGB objects, we calculated the integrated fluxes by
  simply adding up the spectra of each synthetic star.}

\item{We integrated the fluxes {\em analytically} along the
  corresponding isochrones from the MS to the end of central He-burning, using
  a Kroupa IMF \citep{k:01}.}

\item{The choice of the IMF integration constant has been tied to the number of stars
  along the corresponding TP-AGB phase, as follows.  There is a region
  of overlap between the MC simulations and the isochrones, that is
  the E-AGB phase. 
  From each MC simulation of the AGB, we determined the
  number of E-AGB objects, to compare with the number of E-AGB stars predicted by the analytical integration along 
  the isochrone with the Kroupa IMF.
  The IMF integration constant 
  was therefore obtained by imposing that the integration
  along the E-AGB part of the isochrone yields the same number of
  stars of the corresponding MC simulation.
  The integrated flux from the main sequence to the end of the TP-AGB phase 
  was finally computed by 
  summing up the integrated flux from the E-AGB + TP-AGB MC simulation, 
  and the flux from the isochrone integration until the end of central He-burning.  
}

\item{With the IMF integration constant fixed, we determined also the
  total {\sl evolving} mass of the SSP -- stellar mass excluding remnants like white dwarfs,
  neutron stars and black holes -- for each number of AGB stars in the
  MC simulations. Masses below 0.5$M_{\odot}$ and down to
  0.1${\rm M_{\odot}}$, are assumed to contribute only to the total
  mass budget, but not to the fluxes. As a rough guideline, for our two selected metallicities and age ranges, 
  10, 100 and 1000 AGB objects correspond to 
  $\approx 10^4-10^5\, M_{\odot}$, $\approx 10^5-10^6\, M_{\odot}$ and 
  $\approx 10^6-10^7\, M_{\odot}$ of evolving stars, respectively 
  (the exact values depend on age and initial chemical composition).}

\end{itemize}

\begin{figure}
\centering
  \resizebox{16pc}{!}{\includegraphics{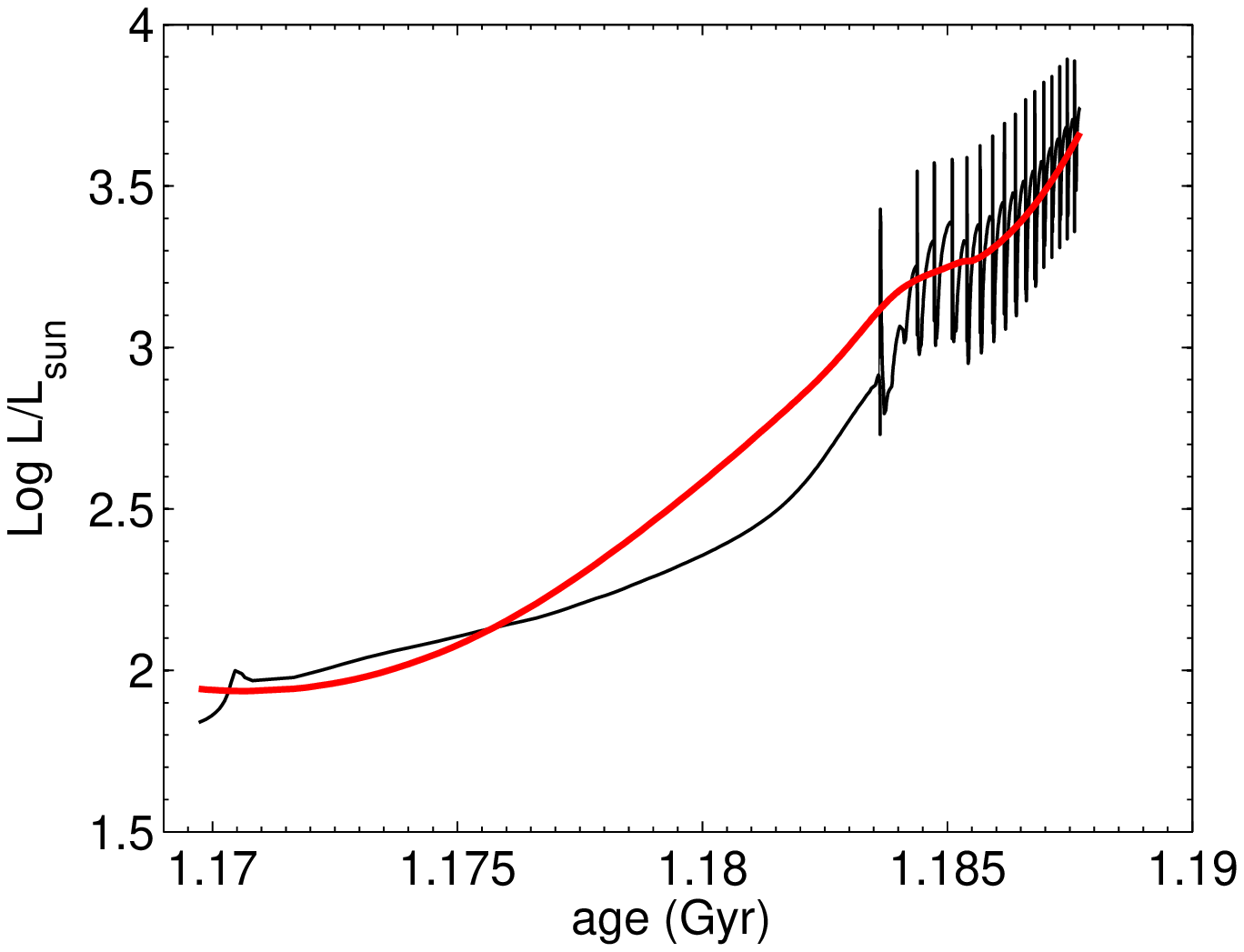}}
  \resizebox{16pc}{!}{\includegraphics{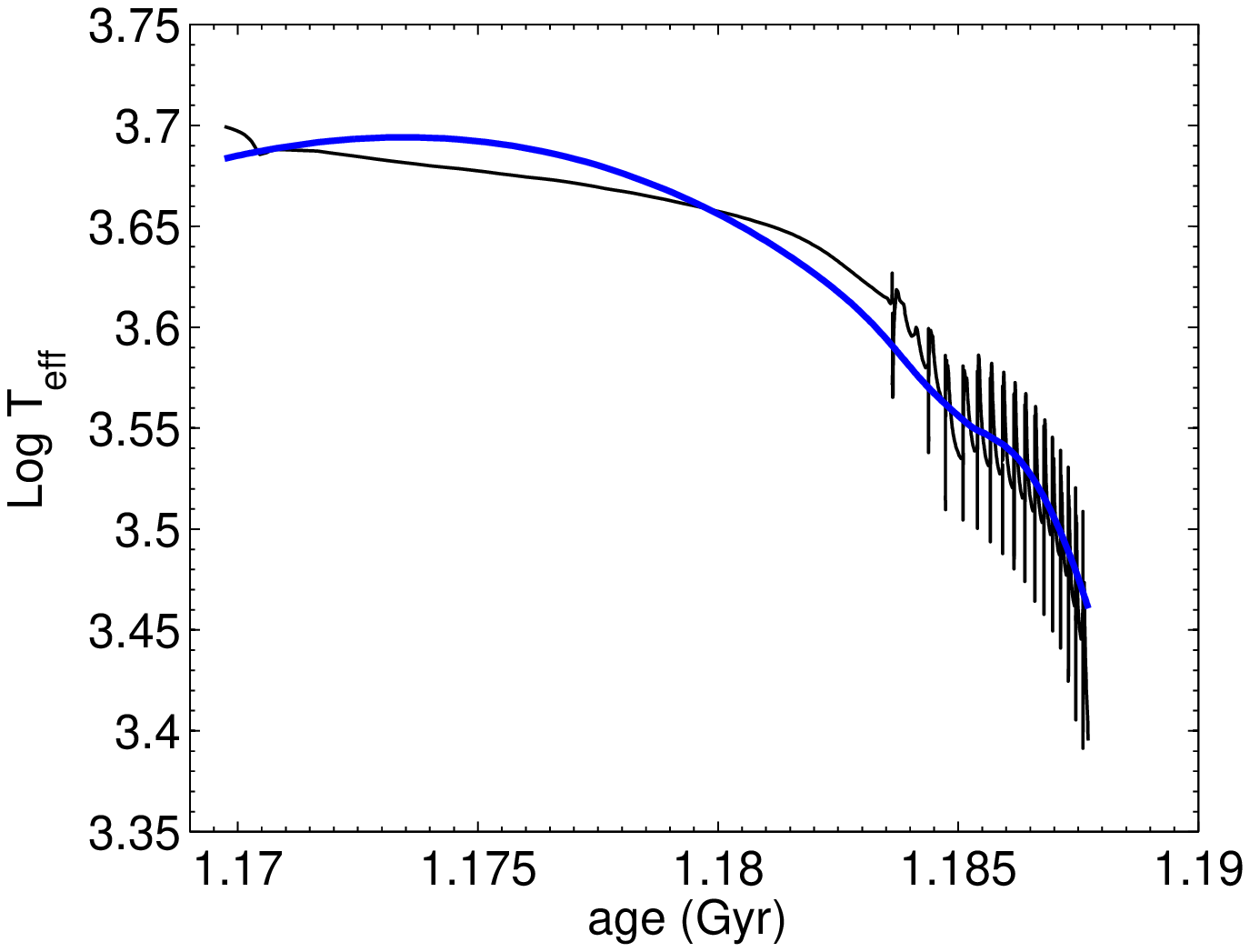}}
\caption{Evolution with age of bolometric luminosity (top) and effective temperature (bottom) of the 
AGB track by WF09 with $Z=0.008$ and M=2$M_{\odot}$ (corresponding to the log(t)=9.08 isochrone), 
and the corresponding smooth version (coloured lines) calculated as in Paper~I.}
\label{smoothtr}
\end{figure}

This whole procedure to calculate integrated fluxes 
was repeated again using the same  
AGB tracks but with the $L$-$T_\mathrm{eff}$ evolution with time 
smoothed as discussed in Sect. 3.2 of Paper~I, and shown in Fig.~\ref{smoothtr}.
These smooth versions of the tracks (that preserve the correct evolutionary timescales 
of the original calculations) are calculated using an analytical fit to the 
original track by employing second order polynomials. Notice how in the example displayed in 
Fig.~\ref{smoothtr}, the smooth track differs from the original one also during the E-AGB phase. 
In other cases the E-AGB evolution is almost unchanged. 
These smooth tracks will be 
employed in the next section to assess quantitatively 
the impact of this procedure on the resulting integrated magnitudes.

\section{Theoretical predictions}
\label{s:prediction}


We begin the presentation of our results by discussing the 
contribution of the AGB to the integrated K-band magnitude -- denoted as $
  L_{K_s}^{AGB}/L_{K_s}^{tot}$ -- of SSPs \citep[we chose the 
2MASS $K_s$-band,][]{skrutskie:97}. This quantity is very  
sensitive to the AGB \citep[see, for example,][]{mbga:2011} and is frequently used as a
standard diagnostics of AGB calculations  
\citep[e.g.][]{mofmt:06, ko:13}. Table~\ref{t:2} summarizes the mean values obtained for 
our MC calculations at $Z=0.004$ and $Z=0.008$. 
We stress again 
that the population ages are determined by the underlying mass for which
the AGB evolution was calculated (Sect.~\ref{s:spectra}), and
therefore are slightly different for the two representative mixtures. 
The spreads around the mean values 
were determined from the $1\sigma$-ranges of the 100 MC
simulations for each case. 

The maximum
contribution of AGB stars to the total K-band luminosity happens
around an age t=1.19~Gyr (log(t)$\sim$9.08) for $Z=0.008$, in agreement with, for example,
the behaviour found by \citet[Fig.~5]{mgba:2010}. 
For the case with $Z=0.004$ the almost constant AGB fraction with age also agrees
with results shown in \cite{mgba:2010}, but smaller variations occuring on short timescales, 
as visible in that paper are no longer discernible because of our
restricted number of isochrone ages. The maximum of the AGB 
contribution to the integrated 
$K_s$-band magnitude from the models is around 1.05~Gyr (log(t)$\sim$9.04).

\begin{figure}
\centering
\includegraphics[width=0.8\columnwidth]{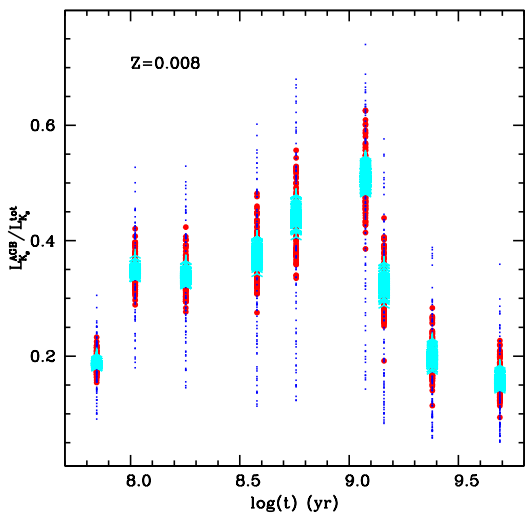}
\caption{Distribution of the $L_{K_s}^{AGB}/L_{K_s}^{tot}$ ratios 
predicted by our MC simulations with   
10, 100, and 1000 AGB stars (symbols size increasing with increasing number of stars), 
metallicity Z=0.008, and the whole set of available ages.} 
\label{f:LkLtotdistr}
\end{figure}

The effect of statistical fluctuations for small numbers of AGB stars is very evident. 
The mean value of $L_{K_s}^{AGB}/L_{K_s}^{tot}$ at fixed age and Z is systematically lower 
for small samples, due to the lack of brighter (and shorter lived) AGB objects, 
and converges to a stable value only when the number of AGB stars reaches $\sim$100. 
The $1\sigma$ dispersion around the mean at fixed age and metallicity 
is obviously also a strong function of the number of AGB stars, 
decreasing with increasing number of AGB objects.
The maximum of the AGB contribution is found at ages around log(t)=9.0-9.1, with a slow rise at younger ages 
followed by a fast drop at older ages.  


Figure~\ref{f:LkLtotdistr} displays graphically the results for the $Z=0.008$ case and the whole 
age range covered by our calculations. It is very evident, especially 
in the simulations for 10 and 100 AGB stars --as also demonstrated by the data in Table~\ref{t:2}--  
how the full range of values 
spanned by the MC realizations at fixed number of AGB objects is a function of age. The total spread 
at fixed age is larger 
at those ages where the mean value of $L_{K_s}^{AGB}/L_{K_s}^{tot}$ is larger, that is, when the AGB contribution 
to the integrated luminosity in the K-band is maximized.
The case with 10 AGB stars is particularly interesting because 10 (or less) 
is the typical number of objects populating the AGB of LMC and SMC clusters employed 
to test the integrated near-IR SSP colours \citep[see, e.g.,][]{mofmt:06,noel:13}. 

\begin{table*}
\caption{Mean values of the fractional AGB contribution to the total luminosity
  in the 2MASS $K_s$-band \citep{skrutskie:97} for our integrated stellar
  populations as a function 
  of age (cols.~2 and 6). The number $N$ of AGB stars in the simulations
  and the two chemical compositions are indicated. Results for the cases
  with 30 and 300 AGB stars are omitted. The  $1\sigma$ spreads are
  determined from the fluctuations in the 100 MC simulations for each case.}
\label{t:2}
\centering
\begin{tabular}{c | c c c c | c c c c }   
\hline\hline
& \multicolumn{4}{c|}{$Z=0.004$} & \multicolumn{4}{c}{$Z=0.008$} \cr
\hline
   & t(Gyr) & N=10 & N=100 & N=1000 ~~~~ & t(Gyr) & N=10 & N=100 & N=1000 \cr
\hline 
$L_{K_s}^{AGB}/L_{K_s}^{tot}$ & 0.07 & 0.255$\pm$0.046 &  0.258$\pm$0.017 & 0.261$\pm$0.005 & 0.07 & 0.188$\pm$0.043 & 
0.189$\pm$0.016 & 0.191$\pm$0.004 \\
 & 0.10 & 0.494$\pm$0.082 &  0.501$\pm$0.028 & 0.506$\pm$0.009 & 0.11 & 0.347$\pm$0.073 & 
0.349$\pm$0.026 & 0.353$\pm$0.008 \\
 & 0.17 & 0.440$\pm$0.086 &  0.439$\pm$0.030 & 0.446$\pm$0.009 & 0.18 & 0.330$\pm$0.088 & 
0.338$\pm$0.031 & 0.343$\pm$0.009 \\
 & 0.35 & 0.397$\pm$0.120 &  0.413$\pm$0.043 & 0.423$\pm$0.012 & 0.38 & 0.345$\pm$0.122 & 
0.373$\pm$0.043 & 0.378$\pm$0.013 \\
 & 0.51 & 0.455$\pm$0.141 &  0.491$\pm$0.047 & 0.501$\pm$0.014 & 0.57 & 0.405$\pm$0.147 & 
0.436$\pm$0.051 & 0.445$\pm$0.015 \\
 & 1.05 & 0.624$\pm$0.137 &  0.662$\pm$0.036 & 0.670$\pm$0.011 & 1.19 & 0.452$\pm$0.147 & 
0.506$\pm$0.049 & 0.512$\pm$0.015 \\
 & 1.30 & 0.406$\pm$0.163 &  0.445$\pm$0.051 & 0.456$\pm$0.016 & 1.44 & 0.280$\pm$0.130 & 
0.321$\pm$0.045 & 0.327$\pm$0.014 \\
 & 2.12 & 0.197$\pm$0.093 &  0.222$\pm$0.034 & 0.226$\pm$0.011 & 2.40 & 0.175$\pm$0.086 & 
0.199$\pm$0.032 & 0.201$\pm$0.011 \\
 & 4.21 & 0.158$\pm$0.067 &  0.175$\pm$0.025 & 0.178$\pm$0.008 & 4.89 & 0.142$\pm$0.069 & 
0.160$\pm$0.028 & 0.161$\pm$0.018 \\
\hline 
\end{tabular}
\end{table*}

\begin{table*}
\caption{As Table~\ref{t:2} but for the integrated$(V-K_s)$ colours 
as a function of age.}
\label{t:3}
\centering
\begin{tabular}{c | c c c c | c c c c }   
\hline\hline
& \multicolumn{4}{c|}{$Z=0.004$} & \multicolumn{4}{c}{$Z=0.008$} \cr
\hline
   & t(Gyr) & N=10 & N=100 & N=1000 ~~~~ & t(Gyr) & N=10 & N=100 & N=1000 \cr
\hline 
$(V-K_s)$ & 0.07 & 1.914$\pm$0.169 & 1.914$\pm$0.058 & 1.921$\pm$0.018 & 0.07 & 2.146$\pm$0.109 & 
2.141$\pm$0.038 & 2.145$\pm$0.012 \\
 & 0.10 & 2.080$\pm$0.179 &  2.083$\pm$0.065 & 2.092$\pm$0.019 & 0.11 & 2.132$\pm$0.123 & 
2.131$\pm$0.045 & 2.136$\pm$0.013 \\
 & 0.17 & 1.906$\pm$0.115 &  1.900$\pm$0.058 & 1.910$\pm$0.017 & 0.18 & 1.874$\pm$0.143 & 
1.880$\pm$0.050 & 1.886$\pm$0.015 \\
 & 0.35 & 1.577$\pm$0.150 &  1.583$\pm$0.078 & 1.597$\pm$0.023 & 0.38 & 1.833$\pm$0.199 & 
1.864$\pm$0.74 & 1.870$\pm$0.022 \\
 & 0.51 & 1.755$\pm$0.195 &  1.773$\pm$0.097 & 1.788$\pm$0.029 & 0.57 & 1.925$\pm$0.267 & 
1.956$\pm$0.097 & 1.969$\pm$0.028 \\
 & 1.05 & 2.442$\pm$0.368 &  2.503$\pm$0.114 & 2.520$\pm$0.036 & 1.19 & 2.341$\pm$0.277 & 
2.421$\pm$0.103 & 2.430$\pm$0.030 \\
 & 1.30 & 2.078$\pm$0.207 &  2.091$\pm$0.098 & 2.107$\pm$0.031 & 1.44 & 2.345$\pm$0.194 & 
2.398$\pm$0.069 & 2.405$\pm$0.023 \\
 & 2.12 & 2.228$\pm$0.123 &  2.258$\pm$0.044 & 2.261$\pm$0.015 & 2.40 & 2.532$\pm$0.115 & 
2.559$\pm$0.042 & 2.560$\pm$0.014 \\
 & 4.21 & 2.400$\pm$0.081 &  2.423$\pm$0.021 & 2.424$\pm$0.010 & 4.89 & 2.711$\pm$0.088 & 
2.731$\pm$0.035 & 2.733$\pm$0.011 \\
\hline 
\end{tabular}
\end{table*}

\begin{figure}
\centering
\includegraphics[width=0.8\columnwidth]{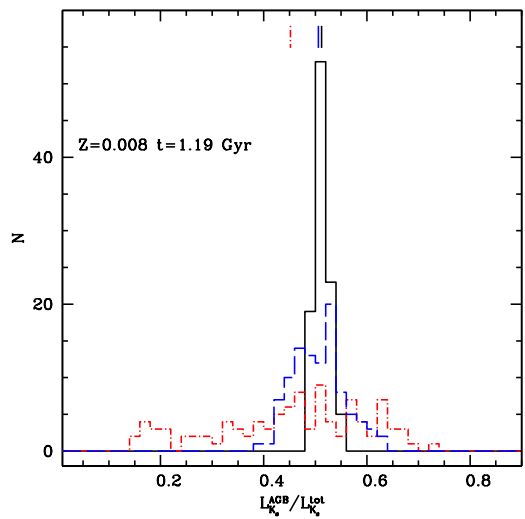}
\caption{Number distribution of $L_{K_s}^{AGB}/L_{K_s}^{tot}$
  values (bin size equal to 0.02) for our MC simulations with   
10 (dot-dashed), 100 (dashed), 1000 (solid) AGB stars, metallicity Z=0.008
including the effect of circumstellar dust, and age t=1.19~Gyr (log(t)=9.08).  
Average values are marked at the top of the diagram (the average values 
for the 100 and 1000 star simulations are essentially the same).} 
\label{f:LkLtot008}
\end{figure}

To delve further into this issue, we show as an example in 
Fig.~\ref{f:LkLtot008} the distribution of 
$L_{K_s}^{AGB}/L_{K_s}^{tot}$ values for $Z=0.008$, an age of 1.19~Gyr (log(t)=9.08, 
when the AGB contribution attains its maximum), and the same 
three values of the total number of AGB stars as in Fig.~\ref{f:LkLtotdistr}.  
It is important to notice that for small numbers of stars the mean
value (that is lower than the case of more populated AGBs) 
does not coincide with a pronounced maximum in the number distribution. 
The probability of encountering a cluster with only 10 AGB stars at any
$L_{K_s}^{AGB}/L_{K_s}^{tot}$ value within a range between $\sim$0.1 and 
and $\sim 0.7$, has an almost flat distribution.
This means that one expect a broad, almost uniform distribution of 
observed $L_{K_s}^{AGB}/L_{K_s}^{tot}$ 
ratios, when comparing models with individual Magellanic Cloud clusters. 
This will be discussed in the next section.

\begin{figure}
\centering
\includegraphics[width=0.7\columnwidth]{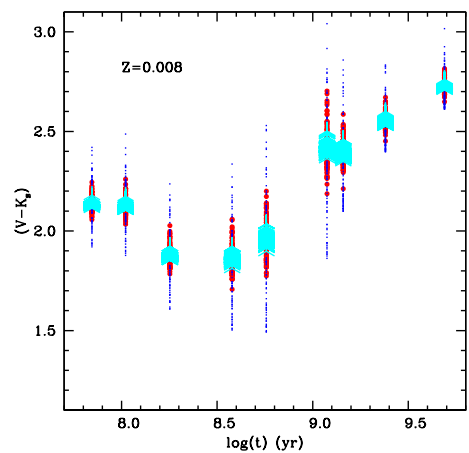}
\caption{The same as Fig.~\ref{f:LkLtotdistr}, but for the $(V-K_s)$ colour.} 
\label{f:VKdistr}
\end{figure}

The spread in $L_{K_s}^{AGB}/L_{K_s}^{tot}$ causes a spread of the 
predicted IR colours \citep[see e.g.,][for past analyses of 
statistical fluctuations on IR colours due to the AGB population]{sf:1997, 
lm:00, b:02}, as demonstrated by the the data in Table~\ref{t:3}, that show 
$(V-K_s)$ colours corresponding to the same 
$L_{K_s}^{AGB}/L_{K_s}^{tot}$ values of Table~\ref{t:2}.   
Figure~\ref{f:VKdistr} displays graphically the 
distribution of $(V-K_s)$ colours from the MC simulations for $Z=0.008$ reported in 
Table~\ref{t:3}. 
Notice how this colour displays a steady increase with age, for ages above 
log(t)$\sim$9.1, after the AGB contribution to the K-band has dropped significantly, due 
to the increasing contribution of RGB stars.
  
\begin{figure}[b]
\centering
\includegraphics[width=0.7\columnwidth]{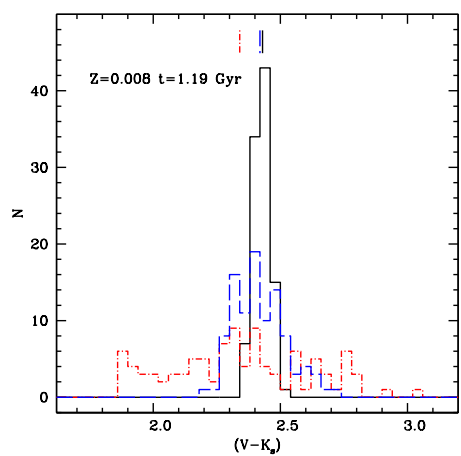}
\caption{As Fig.~\ref{f:LkLtot008} but for the $(V-K_s)$ integrated colour.}
\label{f:VK008}
\end{figure}

The general behaviour of the statistical fluctuations is similar to the case 
of the $L_{K_s}^{AGB}/L_{K_s}^{tot}$ ratio. For the relevant case of 10 AGB objects discussed before, 
the full colour range at the age when the AGB contribution is the largest (log(t)=9.08) 
spans $\approx$1~mag. Figure~\ref{f:VK008} 
displays the distribution of $(V-K_s)$ colours for the same
parameters as in Fig.~\ref{f:LkLtot008}. 
The probability of encountering a cluster with only 10 AGB stars at any
colour between 2.8-3.0 and $\sim$1.9~mag has again an almost flat distribution, and there is 
a blueward bias of $\sim$0.1~mag for the corresponding mean value, compared to the case of more populated AGBs.

\begin{figure}
\centering
\includegraphics[width=0.7\columnwidth]{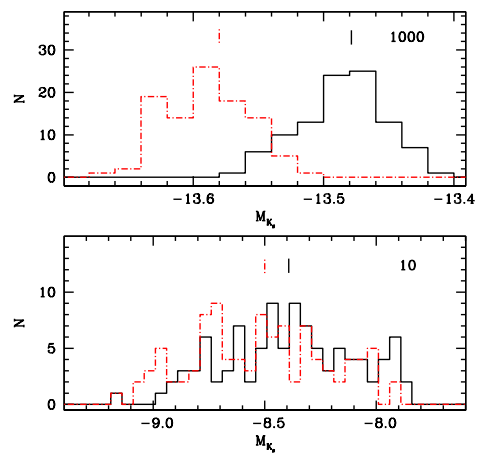}
\caption{Number distribution of integrated ${\rm K_s}$ magnitudes obtained from 
the original (solid line) and smooth (dash-dotted line) AGB tracks, 
for the case with age t=1.19 Gyr (log(t)=9.08) and $Z=0.008$, with 10 (lower panel) 
and 1000 (upper panel) AGB stars. Average values 
are marked at the top of the diagrams.
}
\label{smooth}
\end{figure}

We now compare the results obtained from the full AGB calculations, with the case 
of employing the smooth AGB tracks of Paper~I (see Fig.~\ref{smoothtr}), to assess the presence of 
systematic effects in the contribution of the AGB to the integrated light, due to the smoothing process. 
To this purpose, we calculated sets of integrated magnitudes for $Z=0.004$ and $Z=0.008$ as described in the previous section, 
but employing the smooth version of WF09 AGB tracks.
As expected, blue-optical filters are unaffected by the smoothing of the AGB, whereas integrated magnitudes in 
filters like $JHK$ display some differences, although these are generally minor.
Obviously the larger discrepancies are found at ages where the AGB contribution to the integrated light is larger, e.g. 
around log(t)=9.0. If we focus on the integrated magnitude in the $K_s$-band, the mean values differ by at most $\sim$0.10~mag 
for the cases with log(t)=9.02, $Z=0.004$, and log(t)=9.08, $Z=0.008$,  almost independent 
of the number of AGB stars in the MC simulations. 
In most cases these differences are however within a few 0.01~mag.
Figure~\ref{smooth} displays, as an example, the number distribution of integrated $K_s$ magnitudes obtained from 
the original and smooth AGB tracks, for the case with log(t)=9.08 and $Z=0.008$ (for both 10 and 1000 AGB stars). 
The difference of the mean values is 0.10~mag, smooth tracks providing brighter integrated magnitudes.
In the case with 1000 stars the differences of the overall $K_s$-band magnitude distributions are obvious; as for 
the samples with 10 stars, although the overall distributions seem similar (almost flat over a large 
magnitude range), in addition to the differences of the mean values, a KS-test returns a 99\% probability that 
the two distributions are not the same.

The origin of these differences in the mean values of the integrated magnitudes can be traced 
back not to the TP-AGB phase, 
but to the differences along the E-AGB between the smooth 
analytic approximations of Paper~I and the complete calculations. As shown 
by Fig.~\ref{smoothtr}, the smooth AGB track corresponding to log(t)=9.08 and $Z=0.008$ is on average overluminous 
during the E-AGB phase, and this produces the different mean integrated $K_s$ magnitude.
Smoothing the TP-AGB part of the evolution does not affect appreciably the final integrated magnitudes.

As for the 1$\sigma$ dispersion around the mean magnitudes, it is  
generally only slightly reduced in case of the smooth tracks, for simulations up to 100 AGB stars. 
The reduction of the dispersion is enhanced when the AGB contribution peaks.
To give more quantitative estimates of the maximum expected reduction,
populations of 1000, 100 and 10 
objects display 1$\sigma$ dispersions equal to $\sim$0.03, 0.10 and
0.30~mag for log(t)$\sim$9.0, respectively, but almost independent of age and $Z$.
 These are reduced by $\sim$0.02 and $\sim$0.05-0.10~mag for the case of 
100 and 10 AGB stars respectively, when smooth models are employed.

\begin{figure}
\centering
\includegraphics[width=0.8\columnwidth]{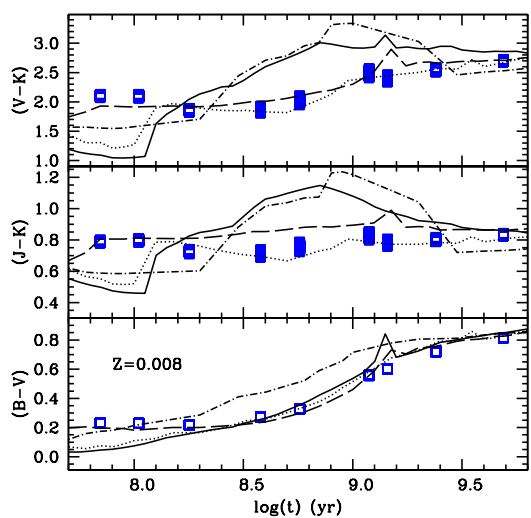}
\caption{Integrated optical and near-IR colours as a function of age for 
$Z=0.008$, as obtained in this paper (open squares -- results for MC simulations with 
1000 AGB stars. Notice the overlap of the symbols 
in the upper two panels, due to small magnitude fluctuations, that mimic  
solid squares), Paper~I (dotted line), BaSTI (dashed line), 
\citet{mar:08} and \citet{mar:05} (solid and dot-dashed, respectively)
population synthesis models (see text for details). 
}
\label{comp008}
\end{figure}

We conclude this section by comparing selected integrated optical and IR colours we 
obtained with WF09 calculations for both AGB and pre-AGB phases 
(we display the full set of MC simulations with 1000 stars for each age), with the results of Paper~I, 
and with the predictions by \citet{mar:05} based on the fuel
consumption theorem formalism, \citet{mar:08},
who made use of sophisticated synthetic AGB calculations, and finally ``BaSTI
 TP-AGB extended'' \citep[the set that includes MS convective core
   overshooting]{bastint}, that employ the synthetic AGB treatment described by \citet{cordier}.
Figure~\ref{comp008} displays the various predictions as a function of age for $Z=0.008$ in the 
Bessel-Brett system. Our colours and Paper~I results in this case do not include the effect of circumstellar dust, 
to be consistent with the other models displayed.
The effect of dust is however completely negligible for the (B-V) colour, and amounting to just 
a few hundredths of a magnitude for (J-K) and (V-K).

We start by considering Paper~I models. We recall that the differences to our results 
involve the pre-AGB evolution, taken 
from independent models, the smoothing of the AGB tracks and their shift in both $T_\mathrm{eff}$ 
and luminosity, to match the 
beginning of the AGB phase to the pre-AGB isochrones. 
The size of these shifts displayed a strong dependence on mass and metallicity of the models (see Fig.~2 in Paper~I).
At this metallicity, the maximum shift in $T_\mathrm{eff}$ is of about 
$\sim$700~K at M=1$M_{\odot}$.
More typical values for the other masses are of the order of 250~K.
As for the luminosity, the shifts are at most $\Delta (L/L_{\odot})\sim$0.1.

In spite of these differences, both optical (unaffected by the AGB) and near-IR colours show a very good 
agreement with our results. Clearly the various effects cancel out to provide such a good agreement 
with fully consistent colours calculated here.
The only major differences appear for ages below 100~Myr, before 
the onset of the AGB phase, our colours being systematically redder.
As for the comparison with BaSTI, also in this case the agreement with our results is 
surprisingly good, even below 100~Myr. There is just a small systematic offset in 
$(J-K)$, for the age range 
dominated by the AGB, BaSTI colours being redder by about 0.05-0.10~mag. 
Notice the moderate red spike of the \citet{mar:08} and BaSTI results at log(t)$\sim$9.1-9.2 
\citep[more evident in $(J-K)$ and $(V-K)$ for BaSTI models, whereas it does not appear in the 
$(J-K)$ colours from][]{mar:08} for the
reasons discussed in the previous section. 

In the case of \citet{mar:05} and \citet{mar:08} models, Fig.~\ref{comp008} 
shows a much larger disagreement with our results. Our near-IR colours are systematically bluer 
between log(t)$\sim$8.3 and log(t)$\sim$9.3 -- where the AGB controls the IR  
luminosity -- by up to $\sim$0.4-0.5~mag in $(J-K)$, and $\sim$1~mag in $(V-K)$.
In the same age range, optical colours display a negligible difference with 
\citet{mar:08} results, whereas \citet{mar:05} are still systematically redder, albeit by a smaller amount.

\subsection{Fluctuations of the RGB integrated luminosity}

So far -- and we will continue to do so in the rest of the paper -- we have focused on the 
statistical fluctuations of the AGB contribution to integrated magnitudes and colours, but also 
the contribution of RGB stars can fluctuate statistically when the number of objects is small. 
This is due to the relatively short evolutionary timescales along the 
bright part of the RGB, specially close to the tip.
To this purpose, we have performed some numerical tests by calculating integrated magnitudes 
with synthetic HRDs also for the 
RGB part of our isochrones, with the total number of RGB objects  
predicted by the analytical integration for a fixed age, chemical composition 
and number of AGB stars.
Multiple MC realizations of both RGB and AGB phases have been performed, to study how the 
fluctuations of the IR integrated magnitudes and colours discussed before are modified by 
the inclusion of statistical effects along the RGB. 

We found that for ages up to log(t)$\sim$9.1 (about 1.2~Gyr) the additional 
fluctuations are negligible compared to
those induced by the AGB, for every cluster mass explored.
This is not suprising, given that the contribution of the RGB phase to the integrated magnitudes is negligible 
in this age range.
For ages above this threshold -- after the RGB transition -- the 
fluctuations due to RGB and AGB become comparable in size, and  
therefore the total 1$\sigma$ fluctuation of the flux will be $\sim$1.4 times the value due only to  
the AGB. This roughly corresponds to the case -- when only  
statistical fluctuations due to AGB stars are considered -- 
of decreasing the number of AGB stars by a factor about 3 compared to the actual value. 
In the next section with comparisons between theoretical predictions and observations, we neglect 
to account for this additional spread coming from RGB stars when considering ages above log(t)$\sim$9.1, because 
its inclusion does not affect the overall agreement --or lack of -- between theory and observations.  

\section{Observational tests}
\label{s:calibrations}

The comparison of Fig.~\ref{comp008} displays appreciable differences of AGB-dominated colours between 
our calculations and some of the independent results displayed there. 
These differences result from the underlying models, and not from
  our new isochrone approach, as was just shown.
This provides an additional motivation 
to address the issue of the consistency of our present calculations with empirical constraints.

We start by comparing the predicted  $L_{K_s}^{AGB}/L_{K_s}^{tot}$ ratios with the 
data by \citet{ko:13}. These authors provided data for about 100 LMC clusters 
in the age range spanned by our models, and assigned an age to each cluster 
primarily by means of isochrone fitting. For about 20 clusters they resorted to literature 
estimates, mainly obtained from integrated colour analysis, and 
assumed $Z=0.008$ for the whole sample. 
Here we will compare the data with theoretical predictions for $Z=0.008$ at ages below log(t)=9.3, 
$Z=0.004$ above log(t)=9.5, and both metallicities for log(t) around 9.3, as discussed in 
the following comparisons with integrated colours.  

\begin{figure}
\centering
\includegraphics[width=0.8\columnwidth]{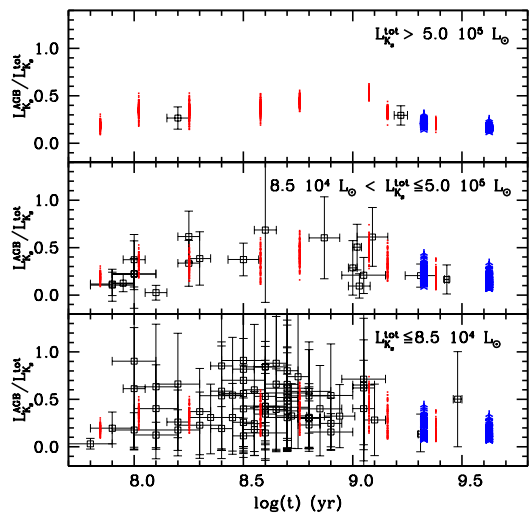}
\caption{$L_{K_s}^{AGB}/L_{K_s}^{tot}$ as a function of age 
  for selected LMC clusters
  (squares with error bars) from \cite{ko:13}, within the displayed 
  ranges of total integrated luminosity. Theoretical predictions for $Z=0.008$ 
  (red three pointed stars) and $Z=0.004$ (blue squares) 
  for the appropriate number of AGB stars are also displayed (see text for details).
}
\label{f:LkLtot_cl}
\end{figure}

Figure~\ref{f:LkLtot_cl} displays the results for the whole sample 
divided into three $L_{K_s}^{tot}$ ranges. Error bars for individual clusters are also displayed, 
taken from \citet{ko:13}.
We compared these data with 
the theoretical predictions from our suite of MC simulations. It is important to notice that 
the K-band integrated luminosity at fixed number of AGB stars has a strong dependence on the age 
(at fixed metallicity).
In the intermediate luminosity range, we selected 
simulations with typically 10 AGB stars, that produce integrated K-band luminosities 
within the selected range 
\citep[in the theoretical calculations we employed $M_{\odot}^{K_s}$=3.27, as in][]{ko:13}. For the 
fainter range we employed also 
the simulations with 10 AGB objects, that is the lowest 
number chosen for our simulations, although they provide integrated
total luminosities still brighter 
that the observed one,  
whilst the two brightest clusters have been compared with the appropriate 
simulations for 30 AGB stars. 

It is very important to notice how the  $L_{K_s}^{AGB}/L_{K_s}^{tot}$ values for the 
faintest sample are scattered almost uniformly over extremely large ranges, 
essentially as predicted for SSPs with a small number of AGB stars 
(see previous section). The intermediate luminosity sample, although of a smaller size, starts to display 
a structure in the  $L_{K_s}^{AGB}/L_{K_s}^{tot}$-age plane that matches the theoretical predictions.
The two most luminous clusters displayed in the top panel, also match very well the theoretical predictions.


To minimize the 
uncertainties of low-number statistics, we followed the approach by
\citet{gonzalez:04}, \citet{Pessev:2008}, \citet{noel:13},
\citet{ko:13}, and 
constructed artificial {\sl superclusters} by adding the fluxes of 
individual, resolved clusters 
which lie in a sufficiently narrow range of
metallicity and age. In general terms, the requirements of this method are obvious:
observations and metallicity determinations should best be from a
homogeneous set of data, and be analysed consistently, and cluster
ages should be reliable. Second, to create a realistic supercluster
this way, both metallicity and age range should ideally be very narrow, which tends to clash 
with the requirement of large AGB star numbers (e.g. several clusters).

\begin{figure}
\centering
\includegraphics[width=0.8\columnwidth]{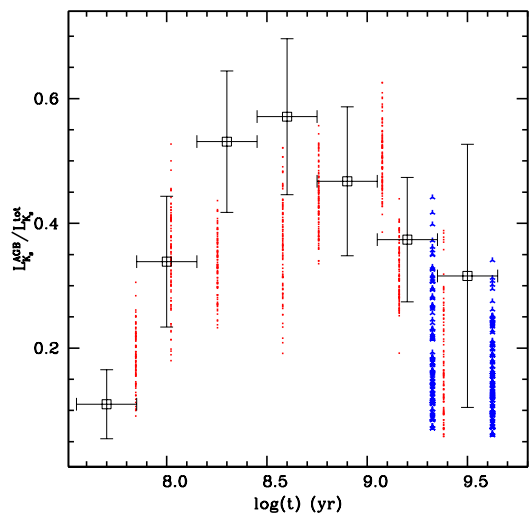}
\caption{As Fig.~\ref{f:LkLtot_cl} but for superclusters  
 obtained from the clusters sample by \cite{ko:13}.} 
\label{f:LkLtot_scl}
\end{figure}

We grouped the clusters from the sample by \cite{ko:13} in 0.3~dex wide age bins 
(the youngest bin centred at log(t)=7.7), and added up both integrated total flux, and the AGB-only flux 
of each cluster in a given group, to produce $L_{K_s}^{AGB}/L_{K_s}^{tot}$ ratios for seven superclusters. \citet{ko:13} calculated also luminosity 
ratios for superclusters, but defined in slightly different age ranges compared to our choice.
Error bars on the resulting $L_{K_s}^{AGB}/L_{K_s}^{tot}$ values have been obtained by a simple error 
propagation of the errors for the individual clusters in each supercluster.
Figure~\ref{f:LkLtot_scl} compares these $L_{K_s}^{AGB}/L_{K_s}^{tot}$  
ratios with theoretical predictions for the number of AGB 
stars that best matches the K-band integrated luminosity of the closest supercluster, 
equal to typically 30 objects (still a reasonably low number). 
The assumption behind the comparison with theory is the following: if theory is in agreement with observations, 
and each supercluster is the equivalent of one theoretical MC realization with the appropriate number of 
AGB stars, the observational points (within their horizontal error bars) 
should overlap with the distribution of theoretical MC simulations at the appropriate age.

It is clear that the data follow very nicely our predictions, and overlap (within the individual error bars) 
with the distribution of MC results. Even if formally the data display a maximum 
at log(t)=8.6, whilst the models predict that the mean value of  $L_{K_s}^{AGB}/L_{K_s}^{tot}$     
has a maximum at log(t)=9.08, there is no real contradiction, given that at log(t)$\sim$8.5 the 
MC simulations predict a range of values that nicely overlap with the observed one. 
  
\begin{figure}
\centering
\includegraphics[width=0.8\columnwidth]{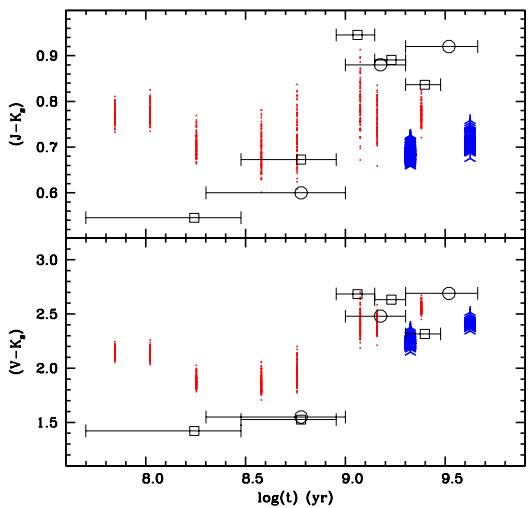}
\caption{Colours of the superclusters (see text for details) by
  \citet{Pessev:2008} (circles with error bars) and \citet{noel:13} 
   (squares with error bars). The horizontal error bars denote 
    the age bins covered by each supercluster. Predictions from
  models for $Z=0.008$ (red dots) and $Z=0.004$ (blue 
  three pointed stars) with a number of AGB stars corresponding to total masses
  a few   times $10^5\, M_{\odot}$ are shown for comparison}
\label{f:supercl}
\end{figure}

As a next step we considered comparisons of selected integrated colours 
with empirical determinations from different sources. 
We start the comparison with {\sl superclusters} from
\citet{Pessev:2008}.
The data were from \citet{pessev:06} for 45
Magellanic Cloud clusters, with an additional 9 clusters taken from the
literature; age estimates come from a variety of sources, mainly based (in the age range of our comparisons) 
on cluster colour-magnitude-diagrams rather than age-sensitive integrated colours. 
Their own 2MASS IR photometry was matched to optical photometry from
other sources. Integrated fluxes were co-added to build several 
superclusters covering age ranges between log(t)=8.3 and the age of
the universe (see Table~5 of that paper), and the 
weighted mean colours with the corresponding errors are provided.
Three of these superclusters
lie in the age range spanned by our models. 
The number of combined clusters (of the order of $\sim$10 per supercluster)   
gives typical masses of the order of a few times
$10^5 M_{\odot}$ for each supercluster. The mean metallicity of each supercluster equals
$\mathrm{[Fe/H]}=-0.34$ for log(t) between 8.3 and 9.0,  
$\mathrm{[Fe/H]}=-0.45$ for log(t) between 9.0 and 9.3, and
$\mathrm{[Fe/H]}=-0.52$ for log(t) between 9.3 and~9.66.  
Integrated colours vs.\ age of these superclusters and our corresponding
theoretical predictions are displayed in  
Fig.~\ref{f:supercl} as open circles. The horizontal error bar correponds to the width of the age bin spanned by each 
supercluster. The error bar on the mean colours has the same size of the symbols.



A similar cluster sample was also
investigated recently by \citet{noel:13}. They adopted  
age estimates for the individual clusters often (but not always) different from \citet{Pessev:2008}, also 
taken from a variety of literature sources, mainly from isochrone fitting. 
These authors also merge clusters into
to superclusters, spanning somewhat narrower age ranges. The typical 
masses estimated by 
the authors are again of order $10^5 M_{\odot}$ for each supercluster, and we plot 
their data in Fig.~\ref{f:supercl} with open squares. The horizontal error bars correspond again 
to the width of the age bin spanned by each supercluster. As for the error on the colours, 
the authors provided the V-band luminosity-weighted standard deviation of the colours in each bin. 
This is different (much larger) from the error due just to the uncertainty on the photometry of individual clusters, 
and we did not display these colour error bars for the reasons we discuss below.


The same Fig.~\ref{f:supercl} shows 
also the predictions from our suite of MC calculations for the number 
of stars that more closely match the 
mass of the closest supercluster (typically we need the results with 100 AGB stars). To follow the trend 
of mean supercluster [Fe/H] with age of \citet{Pessev:2008} data, we employ the results for  
$Z=0.008$, except for the two oldest ages. At ages around log(t)=9.3 we display results for both $Z=0.008$ and $Z=0.004$, 
whereas at the oldest age we show only the $Z=0.004$ results. 
We assume that the trend [Fe/H] vs age for \citet{noel:13} superclusters is consistent with \citet{Pessev:2008}, 
given that they employ a very similar cluster sample. 
Using colours in common to two different sets of superclusters from essentially the same cluster sample, 
that made use of different groupings and age estimates, should give us an idea of the uncertainties involved 
in the empirical determination of the supercluster colours.

The agreement with theory appears satisfactory in the $(J-K_s)$ colour, for the age range between 
log(t)$\sim$8.3 and log(t)$\sim$9.0. At younger ages theoretical colours seem to red, whilst they are 
too blue at older ages compared to \citet{Pessev:2008} data. However, the oldest supercluster  
from \citet{noel:13} does not show a substantial discrepancy with the models for $Z=0.008$.
In case of the$(V-K_s)$ colour, only ages above log(t)$\sim$9.0 are matched satisfactorily by the models; 
at younger ages the models appear systematically too red.

\begin{figure}
\centering
\includegraphics[width=0.8\columnwidth]{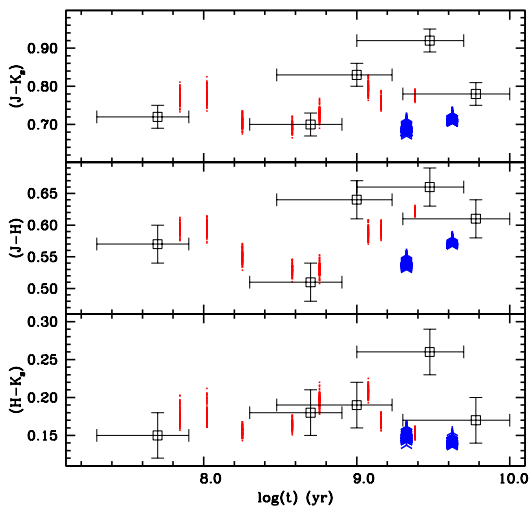}
\caption{As Fig.~\ref{f:supercl}, but for the superclusters by
  \citet{gonzalez:04}. The meaning of symbols is same as in Fig.~\ref{f:supercl}.} 
\label{f:Gonzalezsupercl}
\end{figure}

Figure~\ref{f:Gonzalezsupercl} displays a comparison of theoretical integrated
colours with \citet{gonzalez:04} superclusters, in the age range spanned by our calculations. 
These authors have constructed superclusters from
observed 2MASS integrated magnitudes of about 200 Magellanic Cloud
clusters for which an $s$-parameter \citep{elson:85, elson:88} had
been determined. The SWB-class of each supercluster was defined by the
value of $s$, co-adding the individual clusters of each SWB class
\citep{swb:80}. 
Finally, ages of the various SWB classes were taken
from \citet{cohen:82}. No age estimates of the individual clusters are provided, and   
the mass of the superclusters is usually a factor of about ten higher than the superclusters by  
\citet{noel:13} and \citet{Pessev:2008}.

Also in this case the horizontal error bars shown in the figure corresponds to the width of the age bin spanned by each supercluster. 
As for the colours, we considered the authors' estimate of the photometric errors (0.03~mag).
The theoretical results displayed are from MC calculations with typically 1000 AGB stars (except for the two youngest ages), 
that approximately match the superclusters' masses quoted by \citet{gonzalez:04}.  
The metallicities of the theoretical results 
follow the same trend with age of the previous comparisons \citep[from][superclusters]{Pessev:2008}, that 
is consistent with the metallicity ranges assigned by \citet{gonzalez:04} to each of their superclusters \citep[following][]{cohen:82}.

For these data the agreement with theory is good over the whole age range and the three colour combinations, apart from 
the supercluster centred around log(t)$\sim$9.3. 
Notice that at the youngest ages there is no clear discrepancy between models and observations, 
contrary to what found for \citet{noel:13} and \citet{Pessev:2008} superclusters.
On the other hand, for the supercluster at log(t)$\sim$9.3, the observed colours 
appear redder than theory, especially in the 
$(J-K_s)$ and $(H-K_s)$ colours. The same discrepancy in $(J-K_s)$ is found in the comparison with 
\citet{Pessev:2008} superclusters, but not with \citet{noel:13} ones.
 
It may well be that this situation stems from the effect recently analized in detail by \citet{gmbr:13}.
The basic idea is that for ages at which the RGB starts to develop (transition to electron degenerate He-cores), the 
approximation of constant initial mass for the AGB breaks down \citep[see][for a thorough discussion]{gmbr:13}, and the use 
of our approximation to calculate the AGB evolution underestimates the contribution of these stars to the integrated luminosity 
by up to a factor $\sim$2.
The narrow age range where this happens is at log(t)$\sim$9.2 in our models, which is within the age interval covered by the 
discrepant superclusters. 

To investigate further this issue, we have considered all objects in the \citet{gonzalez:04} supercluster spanning the range 
log(t)=9.0-9.7. 
Most of the clusters are in common with \citet{Pessev:2008} and \citet{noel:13} studies, although in these 
latter works they are grouped into two superclusters rather than one. According to the individual age estimates from these two 
studies, the majority of the objects have an age log(t)$\sim$9.15-9.3, e.g. around the {\sl critical} narrow age range where our approximation 
to calculate integrated colours breaks down.
There is therefore a possibility that the discrepancies seen around log(t)$\sim$9.2-9.3 in Figs.~\ref{f:Gonzalezsupercl} and ~\ref{f:supercl} 
--but that do not show up in Fig.~\ref{f:LkLtot_scl}-- 
might be due to the effect discussed by \citet{gmbr:13}, depending on exact value of the cluster ages assigned 
by the authors, whether clusters around the {\sl critical} age are grouped 
into just one superclusters, and the extent of the 
{\sl dilution} effect due to objects away from the transition to electron degenerate He-cores that are grouped in the same supercluster.

\section{Conclusions}
\label{s:conclusion}

Star clusters in the Magellanic Clouds are of special interest as they
cover an age range in which stars of low- and intermediate mass develop into
double-shell burning objects, populating the AGB, and can be massive enough to host at least a few AGB stars.  
Such clusters therefore serve as template and calibration objects for population
synthesis aimed at understanding the star formation history of distant
galaxies during the last few hundred million to a few billion
years. Observations, such as those shown in Fig.~\ref{f:LkLtot_cl},
however, display a large variation of integrated IR colours for
clusters of presumably very similar age and composition. It is
therefore unclear to what accuracy theoretical population synthesis 
models could be calibrated or verified against such variations.

It was recognized early-on \citep{cbb:1988,fmb:1990} that the low number
of AGB stars in individual clusters results in statistical fluctuations
of the AGB integrated luminosity that, 
given the important contribution of late-type luminous stars
to the integrated IR-light, explains the variation of IR integrated
colours. In this paper we have presented new theoretical predictions of   
near-IR colours of intermediate-age populations, and
tested them against data about clusters in the LMC and SMC, taking into
account their low number of AGB-stars.

Our predictions rest on the new WF09 fully evolutionary 
AGB stellar models for stars of low and
intermediate mass, and a detailed treatment of the
effect of their surrounding dust shells, to achieve the most
realistic description of their stellar energy distribution. 
In Paper~I we have already introduced the
new AGB models and 
the treatment of the dusty envelope, and applied it to
observations of galaxies. In that paper, however, WF09 AGB tracks 
were smoothed out (to calculate AGB isochrones) 
and shifted in both luminosity and $T_\mathrm{eff}$ to match an 
independent set of isochrones that model the pre-AGB evolution.
Here we employ the full evolution until the end of the TP-AGB phase 
from WF09 calculations.

One new aspect of our predictions is the treatment of the TP-AGB phase in
the construction of isochrones (Sect.~\ref{s:spectra}). We have used a MC 
scheme to populate what we call an {\sl evolutionary HRD} with a
given number of AGB stars, comparable to that of observed clusters,
and combine it with a standard isochrone for earlier evolutionary
phases. This way we have been able to take into account all temperature and
luminosity variations that occur during the TP-AGB cycles as well as
the statistical fluctuations due to finite star numbers. Our results,
presented in Sect.~\ref{s:prediction}, quantitatively demonstrate the
possible range of both IR colours (we consider $(V-K_s)$ in
Fig.~\ref{f:VKdistr}) and the contribution of AGB stars to the
integrated $K_s$-luminosity (Fig.~\ref{f:LkLtotdistr}), how they
approach a mean value for large ($\sim 1000$) numbers of AGB stars,
and how they distribute almost uniformly over a wide range for low
numbers (the case of 10 AGB stars). In $(V-K_s)$ the colour of any cluster can vary by
as much as $\pm 0.37$~mag ($1 \sigma$; Table~\ref{t:3}). Our models
cover in fact a range up to $\sim$1.2~mag for ages close to 1~Gyr, when AGB
stars are most dominant. We have also shown that the luminosity and $T_\mathrm{eff}$ variations 
during the TP-AGB cycles do not contribute significantly 
to the total statistical fluctuations of the AGB integrated light, 
such that interpolations amongst smoothed AGB-tracks as in Paper~I, can also be used, when 
carefully constructed. They allow construction of conventional
isochrones for any desired age, while our --in principle-- 
more accurate method is restricted by default to
the ages of the available stellar model tracks.
We have then extended our analysis to investigate the additional contribution of the 
fluctuations of the RGB integrated luminosity for poorly populated clusters. We 
found that, as expected, the RGB contribution is negligible for ages below log(t)$\sim$=9.1 (ages younger than the transition to 
electron degenerate He-cores), while it becomes comparable to the AGB fluctuations at older ages.

In terms of consequences for population synthesis of younger stellar
clusters, e.g. in star forming regions of distant galaxies, our results
imply that for any given age in the range between $\sim$0.1 and $\sim$1 Gyr, infrared
colours may vary --due to statistical fluctuations in the number of
AGB stars-- to an extent that the cluster age cannot be determined more
accurately than a factor $\sim$10. This is already implied by the observational
data used in this paper, but substantiated now by our theoretical models.



We have taken into account these AGB statistical fluctuations to compare our 
predictions with empirical data on Magellanic Cloud clusters, the standard bench tests 
of near-IR population synthesis models.  
The comparison with empirical estimates of the $L_{K_s}^{AGB}/L_{K_s}^{tot}$ ratio 
by \cite{ko:13} shows very good agreement in general: when considering the 
appropriate number of AGB stars, the statistical
scatter of our predictions covers the observed distribution, and the variation of
$K_s$-band luminosity with age for stacked clusters (or {\em
  supercluster}) also agrees very well with their data
(Figs.~\ref{f:LkLtot_cl} and \ref{f:LkLtot_scl}). 

When considering integrated near-IR colours, the level of agreement 
with empirical data for several samples of superclusters 
\citep{noel:13,Pessev:2008,gonzalez:04} depends somehow on the 
selected colour, but also on the
sample. Clearly, the construction of superclusters as well as the
individual observations themselves, are sometimes not consistent. 
These discrepancies need clarification, and should be a
warning to population synthesis models that require calibration with
such observed superclusters. In general, our predictions pass the
observational tests well, and can also be properly applied 
to study individual clusters hosting a low number of AGB stars.

A comparison with integrated colours from 
other population synthesis models, including the results  
of Paper~I, reveals significant discrepancies with results from 
\citet{mar:08} and \citet{mar:05}. \citet{noel:13} have already discussed how these 
models disagree (they are generally too red) with the integrated colours of their superclusters 
for ages between $\sim 10^8$ and $\sim 10^9$~years.
The agreement with the model of
Paper~I and with the BaSTI-model by \citet{cordier} is, in contrast,
excellent. 

In summary, we have presented a statistical approach to model integrated
colours of intermediate-age populations with a small number of stars
that makes use of new state-of-the-art stellar evolution models, the effect of dusty
envelopes, quantifies statistical fluctuations and matches well
observations of individual Magellanic Cloud clusters as well as
artificial superclusters. The predictive power of theoretical models,
which are not calibrated beforehand on a specific set of observations --this calibration being
difficult because of the statistical effects-- should be
sufficient for wider applications.

\begin{acknowledgements}
We thank N.E. N\"oel for discussion about her results, and S. Cassisi for comments and an earlier draft of the paper.
M.S. is grateful to the Max Planck Institute f\"ur Astrophysik for hospitality and support 
for several visits, during which most of this work was carried out.   
\end{acknowledgements}

\bibliographystyle{aa}

\end{document}